\documentclass[english,aps,prb,twocolumn,footinbib,showpacs,showkeys,longbibliography,secnumarabic,superscriptaddress,floatfix]{revtex4}

\usepackage[english]{babel}
\usepackage{amssymb,amsmath,bm,soul}

\usepackage[pdftex]{graphicx}
\usepackage[caption=false,listofformat=subsimple]{subfig}

\usepackage{color}
\usepackage[usenames,dvipsnames]{xcolor}
\usepackage{hyperref} 
\hypersetup{%
	linktoc=page,%
	colorlinks=true,%
	linkcolor=RoyalBlue,%
	citecolor=OliveGreen,%
	urlcolor=OliveGreen,%
	linkbordercolor=red,%
	pdfborderstyle={/S/U/W 1},%
}
\setul{1.4pt}{.4pt}
\newcommand{\uhref}[2]{{\href{#1}{\ul{#2}}}}

\usepackage{enumerate}
\usepackage{grffile}
\usepackage{balance}
\usepackage{float}
\usepackage{listings}

\newcommand{\Jc}{J_\mathrm{c}}
\newcommand{\Ec}{E_\mathrm{c}}
\newcommand{\Hct}{H_\mathrm{c2}}
\newcommand{\Bm}{B_{\scriptscriptstyle \Phi}}
\newcommand{\Jdp}{J_\mathrm{dp}} 
\newcommand{\Jext}{J_{\mathrm{ext},x}}

\newcommand{\Tc}{T_\mathrm{c}}
\newcommand{\Tci}{T_\mathrm{c}^\star}
\newcommand{\ei}{\varepsilon^\star}

\begin{document}

\title{Effect of hexagonal patterned arrays and defect geometry on the critical current of superconducting films}

\author{I. A. Sadovskyy}
\affiliation{Materials Science Division, Argonne National Laboratory, 9700 S. Cass Av., Argonne, IL 60637, USA}
\affiliation{Computation Institute, University of Chicago, 5735 S. Ellis Av., Chicago, IL 60637, USA}

\author{Y. L. Wang}
\affiliation{Materials Science Division, Argonne National Laboratory, 9700 S. Cass Av., Argonne, IL 60637, USA}
\affiliation{Department of Physics, University of Notre Dame, Notre Dame, IN 46556, USA}

\author{Z.-L. Xiao}
\affiliation{Materials Science Division, Argonne National Laboratory, 9700 S. Cass Av., Argonne, IL 60637, USA}
\affiliation{Department of Physics, Northern Illinois University, DeKalb, IL 60115, USA}

\author{W.-K. Kwok}
\affiliation{Materials Science Division, Argonne National Laboratory, 9700 S. Cass Av., Argonne, IL 60637, USA}

\author{A. Glatz}
\affiliation{Materials Science Division, Argonne National Laboratory, 9700 S. Cass Av., Argonne, IL 60637, USA}
\affiliation{Department of Physics, Northern Illinois University, DeKalb, IL 60115, USA}

\date{\today}

\begin{abstract}
Understanding the effect of pinning on the vortex dynamics in superconductors is a key factor towards controlling critical current values. Large-scale simulations of vortex dynamics can provide a rational approach to achieve this goal. Here, we use the time-dependent Ginzburg-Landau equations to study thin superconducting films with artificially created pinning centers arranged periodically in hexagonal lattices. We calculate the critical current density for various geometries of the pinning centers~--- varying their size, strength, and density. Furthermore, we shed light upon the influence of pattern distortion on the magnetic-field-dependent critical current. We compare our result directly with available experimental measurements on patterned molybdenum-germanium films, obtaining good agreement. Our results give important systematic insights into the mechanisms of pinning in these artificial pinning landscapes and open a path for tailoring superconducting films with desired critical current behavior.
\end{abstract}

\pacs{
	74.20.De,		
	74.25.Sv,		
	74.25.Wx,		
	05.10.$-$a	
}

\keywords{
	time-dependent Ginzburg-Landau, 
	large-scale simulations,
	vortex dynamics,
	MoGe,
	patterining
}

\maketitle

\tableofcontents

\section{Introduction}

Superconducting films are exemplary systems to study the effect of different pinning landscapes on the resulting critical current of the system. In general, pinning centers reduce the mobility of magnetic vortices and, as a result, their dissipative effects.\cite{Blatter:1994,Nattermann:2000,Blatter:2003} Experimentally, defects can be introduced into superconducting films in a controllable fashion using advanced nanofabrication techniques such as focus ion beam (FIB) milling,\cite{Latimer:2013} electron beam lithography (EBL),\cite{Kemmler:2016} EBL combined with reactive ion etching,\cite{Wang:2013,Wang:2016} or ion irradiation.\cite{Trastoy:2014} Many artificial pinning array structures (pinscapes) have been studied experimentally as well as numerically, for example, square arrays of antidots,\cite{Baert:1995,Harada:1996,Moshchalkov:1998,Berdiyorov:2006,Berdiyorov:2006b,Berdiyorov:2006c,Berdiyorov:2007,Sabatino:2010,Silhanek:2010,Wang:2016} hexagonal (or triangular) pinning lattices,\cite{Kemmler:2009,Rablen:2011} honeycomb arrays,\cite{Wu:2005,Reichhardt:2007,Reichhardt:2008,Latimer:2012} Penrose lattice arrays,\cite{Misko:2005,Misko:2006,Misko:2010,Kramer:2009,Silhanek:2006,Misko:2012,Rablen:2011} blind hole arrays,\cite{Bezryadin:1996,Raedts:2004,Berdiyorov:2009} diluted periodic arrays,\cite{Guenon:2013,Kemmler:2009} composite lattices,\cite{Silhanek:2005} pinscapes with density gradients,\cite{Wu:2007,Motta:2013,Wang:2013,Guenon:2013,Misko:2012,Ray:2013,Ray:2014,Wang:2016} and pinscapes with geometrically frustrated energy landscape.\cite{Libal:2009,Latimer:2013,Trastoy:2014} These studies were limited to either small sizes, two-dimensional systems, and/or one particular configuration. 

In the present work, we perform a systematic study of various hexagonal pinning lattices using large-scale simulations of the time-dependent Ginzburg-Landau (TDGL) equations. This method can capture the vortex dynamics in a realistic way and is the best compromise between microscopic simulations and describing the phenomenological behavior of a superconducting material using currently available supercomputers.\cite{Sadovskyy:2016a,Koshelev:2016} Within the TDGL approach, the collective dynamics of vortices in thin, but finite-thickness, superconducting films are taken into account automatically and allow us to obtain results on experimentally relevant length scales. We vary the size, strength, and density of the pinning sides and study the underlying vortex dynamics, which explains the magnetic-field-dependent critical current. 

Naively, one can expect to find simple matching effects for the field-dependent critical current, e.g., peaks at field values corresponding to vortex numbers, which are integer multiples of the number of pinning sites (and minor peaks at simple fractions). However, in reality, vortex dynamics and pinning are much more diverse, such that this theoretical picture is not (always) observed. For example, vortices are not only confined to the pinning centers, but also can be caged in between them due to repulsion at certain filling fractions, depending on the geometry of the pinscape.

A second main aspect of this work is to study the influence of distortions on the periodic pinning array, i.e., the question on how matching effects can be destroyed.

Finally, we compare our numerical results with experimental data on artificially patterned molybdenum-germanium (MoGe) films to explain the vortex behavior of these systems by large-scale simulations. This represents another aspect of the \textit{critical-current-by-design} paradigm.\cite{Sadovskyy:2016b}

The article is organized as follows. We introduce the numerical description in Secs.~\ref{sec:model_GL}--\ref{sec:model_criterion} and the experimental technique in Sec.~\ref{sec:experimental_technique}. The comparison between numerical and experimental results is presented in Sec.~\ref{sec:results_comparison}. The dependence on inclusion strength is discussed in Sec.~\ref{sec:results_strength} and the influence of distortions of the periodic pinning arrays is discussed in Sec.~\ref{sec:results_randomness}. We summarize our results in Sec.~\ref{sec:conclusions}.

\begin{figure}
	\begin{center}
		\subfloat{\includegraphics[width=8.6cm]{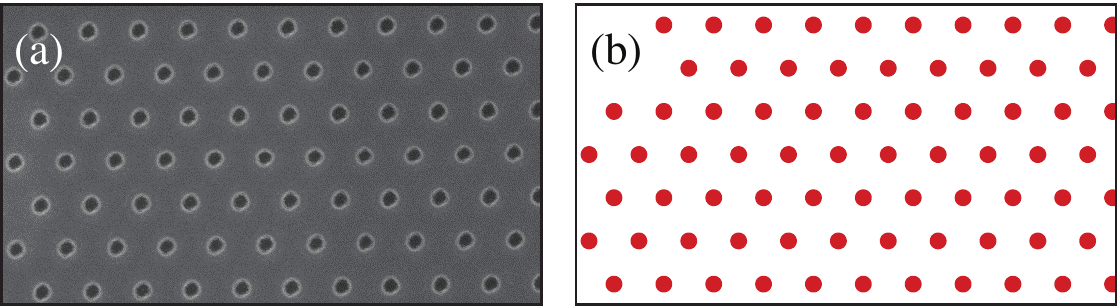} \label{fig:hexagonal_exp}}
		\subfloat{\label{fig:hexagonal_sim}} \vspace{-1.2mm} \newline
		\subfloat{\hspace{-0.25cm} \includegraphics[width=8.9cm]{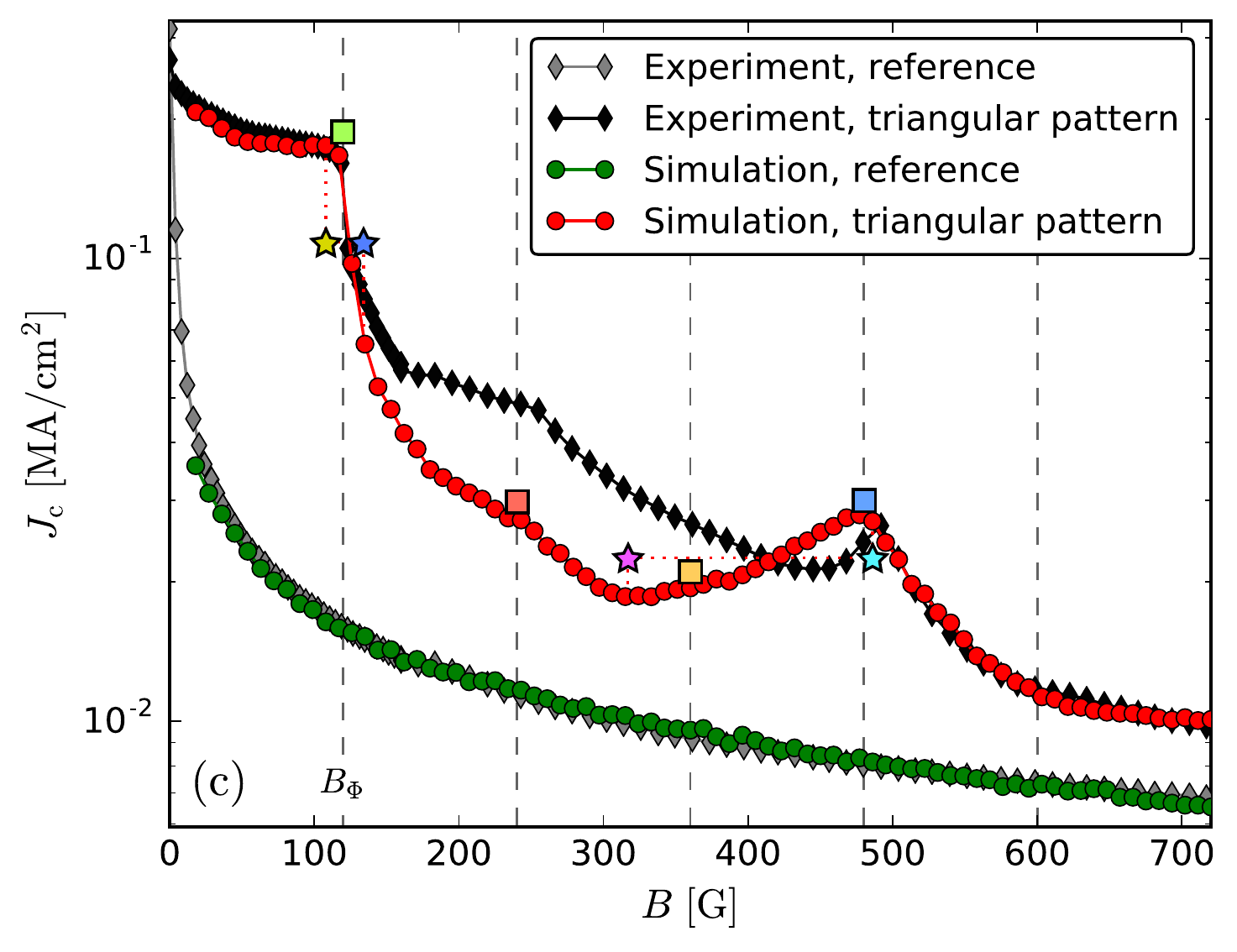} \label{fig:Jc_B_exp_sim}} \vspace{-2mm}
	\end{center} \vspace{-4mm}
	\caption{
		Superconducting film patterned with a hexagonal lattice of defects.
		\protect\subref{fig:hexagonal_exp}.~Scanning electron microscopy image 
		of experimental pattern in a MoGe sample with hole size $d = 150$\,nm 
		and lattice constant $l = 446$\,nm.
		\protect\subref{fig:hexagonal_sim}.~The corresponding $\Tc$ modulation used 
		in the time-dependent Ginzburg-Landau simulations. The (red) dots correspond 
		to the regions with reduced $\Tci$ compared to bulk $\Tc$ in the rest 
		of the superconductor, $\Tci \leqslant \Tc$.
		\protect\subref{fig:Jc_B_exp_sim}.~Comparison of the experimental and 
		numerical $\Jc(B)$ dependence. 
		The experimental reference curve $\Jc(B)$ (gray diamonds) is measured 
		in a pristine MoGe film. In order to match the experimental critical current 
		with the simulation critical current, a small amount of background inclusions 
		was added to the clean simulated system (green circles).
		The addition of a periodic hexagonal array of circular strong pinning centers 
		makes the $\Jc(B)$ dependence nonmonotonic; see, e.g., features 
		in the experimental (black diamonds) curve at $B = \Bm$, $2\Bm$, 
		and $4\Bm$. Here the matching field $\Bm \equiv \Phi_0/S \approx 120$\,G 
		(one inclusion per area $S = \sqrt{3} l^2 / 2$) is shown by vertical dashed lines.
		In particular, the simulations (red circles) reproduce the drop at $B = \Bm$ 
		and the peak at $4\Bm$. However, the behavior between $\Bm$ 
		and $\sim 4\Bm$ is somewhat different. Snapshots of the order parameter 
		at the magnetic field and current values marked by the colored stars 
		and squares in panel~\protect\subref{fig:Jc_B_exp_sim} are presented 
		in Figs.~\ref{fig:psi_sameJ} and \ref{fig:psi_dJ}, respectively.
	}
\end{figure}

\section{Model and Experiments} \label{sec:model}

Here we introduce our model of artificially patterned superconducting films and explain the experimental setup and fabrication of the samples, which we compare with the numerical results.

\subsection{Simulations of patterned superconducting films} \label{sec:model_GL}

In order to model the superconducting films, we use the TDGL equation and apply it to systems with periodically modulated defects.

The TDGL equations effectively capture the collective vortex dynamics and pinning in realistic systems. As the London penetration depth is typically large compared to the coherence length in superconducting films, the TDGL equations can be simplified to just the time evolution of the superconducting order parameter, $\psi = \psi(\mathbf{r}, t)$, with a constant magnetic field $B$ perpendicular to the film, i.e.\cite{Sadovskyy:2015a},
\begin{equation} 
	(\partial_t + i \mu)\psi 
	= \epsilon (\mathbf{r}) \psi - |\psi |^2 \psi + 
	(\boldsymbol{\nabla} - i \mathbf{A})^2 \psi + \zeta(\mathbf{r}, t),
	\label{eq:GL} 
\end{equation} 
where $\mu = \mu(\mathbf{r}, t)$ is the scalar potential, $\mathbf{A} = [0, xB, 0]$ is the vector potential associated with the external magnetic field, and $\zeta(\mathbf{r}, t)$ is an additive thermal-noise term. This equation provides an adequate, quantitative description of strong {type-II} superconductors in the vortex phase. Equation~\eqref{eq:GL} is written in dimensionless units, where the unit of length is the superconducting coherence length~$\xi$, the unit of time is $t_0 \equiv 4\pi \sigma \lambda^2 / c^2$, $\lambda$ is the London penetration depth, $\sigma$ is the normal-state conductance, and the unit of the magnetic field is given by the upper critical field $\Hct = \hbar c / 2e \xi^2$ ($-e$ is the electron's charge and $c$ is the speed of light).

Thermal fluctuations, described by $\zeta(\mathbf{r}, t)$, are determined by its time and space correlations and absolute temperature.\cite{Sadovskyy:2015a} Since we are targeting low temperature vortex behavior, this term becomes negligible. For comparison, the typical temperature is $T \sim 5$\,K in the experiment on MoGe films (see below).

We model the superconducting film as a thin slab with finite thickness. In order to capture the relevant collective vortex dynamics in the system, we simulate a sample of size $512\xi \times 512\xi \times 2\xi = 12.8\,\mu\mathrm{m} \times 12.8\,\mu\mathrm{m} \times 50\,\mathrm{nm}$, where we used the coherence length $\xi(4.8\,\mathrm{K}) \approx 25$\,nm of MoGe for the physical dimensions. This sample is spatially discretized on a regular mesh with $1024 \times 1024 \times 4$ grid points with quasiperiodic boundary conditions in $x$ (along the applied current) and $y$ directions, and open boundary condition in the $z$ direction (along the magnetic field). The thickness of $50$\,nm corresponds to the actual experimental film, but its lateral dimensions are about four times smaller than the actual film size, but are sufficient to capture ``bulk'' properties correctly.

\subsection{Modeling of inherent and artificial defects} \label{sec:model_inclusions}

We model material defects in the Ginzburg-Landau formalism using so-called $\delta \Tc$ pinning, where the critical temperature is spatially modulated due to defects that cause pair-breaking scattering.\cite{Kwok:2016,Berdiyorov:2012a,Berdiyorov:2012b} We use the dimensionless coefficient of the linear term in Eq.~\eqref{eq:GL}, $\epsilon(\mathbf{r}) \propto \Tc(\mathbf{r}) - T$, to introduce spatial $\Tc$ modulations. This means that for $\Tc(\mathbf{r}) < T$, the linear coefficient is negative, which models normal defects, while for $\Tc(\mathbf{r}) > T$ and different from the bulk $\Tc$, weak superconducting defects are captured. We use this modulation to both model the inherent defects of the material as well as the larger-scale artificial periodic pinning array. Our equation is scaled such that $\epsilon(\mathbf{r}) = 1$ for the bulk superconductor (due to the choice of the length scale), i.e., $\epsilon(\mathbf{r}) = \ei = (\Tci - T) / (\Tc - T)$, where $\Tci$ is the critical temperature inside the inclusions and $\Tc$ is the bulk critical temperature. The pinning centers of the array are modeled by short cylinders of diameter~$d$ and height~$h$. Together with the value of $\ei$, these are the parameters controlling the pinning properties of the defects.

Since even the pristine material (without pinning array) typically has a finite critical current, some mechanism of pinning has to be present. In the case of amorphous MoGe films, there are several possible vortex pinning candidates. While these films are typically very flat and homogeneous, they still have some small surface roughness (of the order of a few nanometers) and spatial composition variations. However, it is not known which type of defects causes the inherent pinning. Nevertheless, we need to take these weak pinning centers into account to obtain a finite critical current for the simulated pristine sample. We chose different types of~$\Tc$ modulations in order to match the simulated critical currents with that of the experimental sample without periodic pinning array to obtain a baseline for the pristine sample. For all studied cases~--- weak random modulation near $\epsilon = 1$ everywhere, small polycrystalline Voronoi patterns with typical size of a few coherence length, or isolated small spherical inclusions~--- we could fit the pristine field dependence of the critical current. We therefore chose, throughout this work, to model the inherent inhomogeneities of MoGe by small randomly placed background inclusions of diameter $1.5\xi = 37.5$\,nm with $\ei = 0$ (i.e., $\Tci = T = 5$\,K) occupying 1.8\% of the sample volume.

In contrast to layered high-$\Tc$ materials, MoGe films have no large-scale $\delta\ell$ defects due to the anisotropy of the material, which we therefore did not include in our model.

\subsection{Critical current calculation} \label{sec:model_criterion}

In order to obtain the critical current density from the simulations, we apply a current to the system along the $x$ direction. The total (normal and superconducting) in-plane current density is then given by the expression 
\begin{equation}
	\mathbf{J} 
	= \mathrm{Im} \bigl[ \psi^*(\boldsymbol{\nabla} - i \mathbf{A}) \psi \bigr]
	- \boldsymbol{\nabla} \mu,
	\label{eq:J} 
\end{equation}
in units of $J_0 = \hbar c^2 / 8\pi e \lambda^2 \xi$. For an applied current density $\Jext$, we need to solve an additional ordinary differential equation for the voltage: $\Jext = \langle J_x \rangle_\mathbf{r}$, where $\langle \cdot \rangle_\mathbf{r}$ is the spatial average over the complete system. (The maximum theoretical depairing current density is $\Jdp = 2 J_0 / 3\sqrt{3}$.) In this case, we also need to take into account the condition $\boldsymbol{\nabla} \mathbf{J} = 0$, resulting in the Poisson equation,
\begin{equation}
	\Delta\mu =
	\boldsymbol{\nabla} \,
	\mathrm{Im} [ \psi^*(\boldsymbol{\nabla} - i \mathbf{A}) \psi ],
	\label{eq:Poisson}
\end{equation}
for the scalar potential~$\mu$ in addition to Eq.~\eqref{eq:GL}.

In order to determine the critical current value $\Jc$, we dynamically adjust the external current $\Jext(t)$ in the simulation such that the electric field or voltage drop across the sample has a predefined value. Choosing this electrical field $\Ec$ sufficiently small, and averaging the current density $\Jext(t)$ over different background defect configurations $\mathcal{D}$ and time, we obtain the critical current density $\Jc = \langle \Jext\rangle_{\mathcal{D}, t}$, where $\langle \cdot \rangle_\mathcal{D}$ and $\langle \cdot \rangle_t$ is the average over realizations and time, respectively. Here, we use the finite electric field criterion $\Ec = 10^{-5} E_0$ to determine the critical current, where $E_0 = J_0 / \sigma$ is the electric field unit. Note that the $\Ec$ value used in the numerical simulations is much higher than the level of dissipation corresponding to the value of $2\,\mu$V/cm, routinely used as a practical criterion for $\Jc$ in experiments. Therefore, the simulated critical currents are expected to be somewhat higher than the experimental ones, but due to the large exponent of the $I$-$V$ curve near the transition, there is no qualitative difference expected.

We average the dynamically adjusted current over $10^5$ Ginzburg-Landau time steps and over $|\mathcal{D}| = 16$ different realizations of the initial conditions of the order parameter and random background inclusions. This method gives the same result (within 2\% accuracy) as calculating the current-voltage characteristics at many different applied currents and defining~$\Jc$ from $E = \Ec$ in the $I$-$V$ curve.\cite{Sadovskyy:2016a,Sadovskyy:2016b}

\subsection{MoGe sample and patterning technique} \label{sec:experimental_technique}

Since we compare experimental results on Mo$_{0.79}$Ge$_{0.21}$ films with our simulations, we briefly characterize these samples here. The 50-nm-thick MoGe films were divided into $50\,\mu\mathrm{m} \times 50\,\mu\mathrm{m}$ sections using photolithography and magnetron sputtering deposition. The resulting sample has a critical temperature of $\Tc = 5$\,K, while transport ($I$-$V$) measurements were carried out at temperature $T = 4.8$\,K. We estimate the coherence length in this sample to be $\xi(T) = 25$\,nm, with an upper critical field of $\Hct(T) = \Phi_0/2\pi\xi^2(T) = 5300$\,G, and depairing current density $\Jdp(T) = 0.8$\,MA/cm$^2$ using typical material parameters for MoGe. The hexagonal array of pinning centers of diameter $150$\,nm and lattice constant $446$\,nm was patterned using an EBL patterned nanohole mask and reactive ion etching technique, shown in Fig.~\ref{fig:hexagonal_exp}.

A representation of the $\Tc$ modulation in the simulation for this sample is shown in Fig.~\ref{fig:hexagonal_sim}, where the defects have a diameter of $d = 6\xi$ and $\ei = -24$ and the lattice constant of $l \sim 18\xi$ (in dimensionless units).

\section{Results} \label{sec:results}

\subsection{Simulations vs experiments} \label{sec:results_comparison}

Figure~\ref{fig:Jc_B_exp_sim} demonstrates the comparison between the experimental and numerical $\Jc(B)$ dependence. As mentioned before, we add a small spherical background defect to the clean simulation system with diameter $37.5$\,nm, occupying 1.8\% of the sample volume, in order to reproduce the experimental reference curve (gray diamonds) of the pristine sample. The resulting numerical $\Jc(B)$ dependence averaged over 16 realizations of disorder is shown by green circles and nicely coincides with the experimental reference curve shown by gray diamonds.

The red curve in Fig.~\ref{fig:Jc_B_exp_sim} demonstrates the simulated $\Jc(B)$ dependence in the presence of a hexagonal lattice of patterned defects. For the simulations, the diameter of the holes is $d = 150$\,nm and the lattice constant is $l = 446$\,nm, mimicking the experiment. We chose the depth of the inclusions to be $h = 12$\,nm and the critical temperature inside the inclusions as $\Tci = 0$. The resulting simulated curve coincides well with the experimental one at fields~$B$ lower than $\Bm$ and higher than $4\Bm$. Field values that are integer multiples of the matching field $\Bm \equiv \Phi_0/S = 120$\,G [the field generating a single vortex per inclusion or per lattice unit-cell area $S = \sqrt{3} l^2 / 2 = (415\,\mathrm{nm})^2$] are shown by vertical dashed lines. For $\Bm < B < 4\Bm$, the simulations show somewhat smaller critical currents compared to the experiment. However, simulations reproduce almost all of the qualitative features of the curve such as the minimum at $B = \Bm$ and maximum at $4\Bm$. At $\Bm$, we see a down kink resulting from the vortex caging effect, where an additional vortex is pinned in the center of a hexagonal vortex lattice cell by repulsive forces from vortices that reside in the inclusions.\cite{Berdiyorov:2006b} 

\begin{figure}
	\begin{center}
	\subfloat{\href{https://youtu.be/OOzHO7OTqSg}{\includegraphics[width=8.7cm]{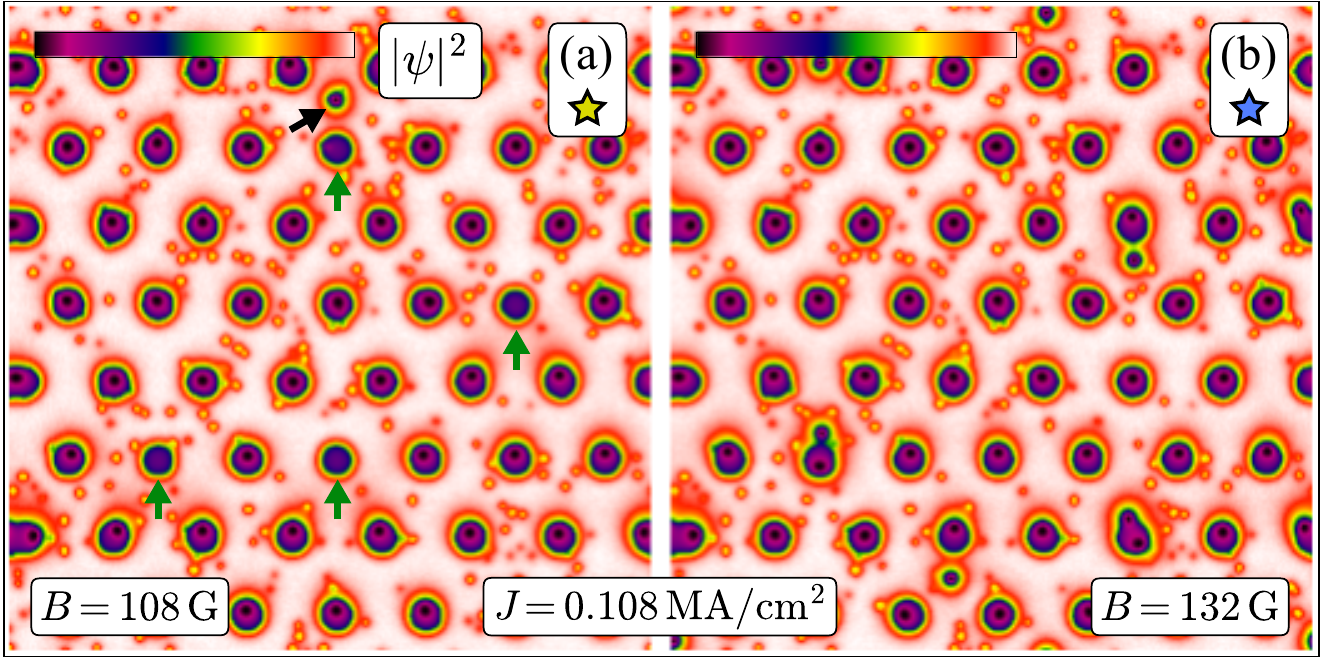}}\label{fig:psi_h12nm_B108G_J0p108MAcm2}}
	\subfloat{\label{fig:psi_h12nm_132G_J0p108MAcm2}} \newline
	\subfloat{\href{https://youtu.be/TAfHOG_lfEk}{\includegraphics[width=8.7cm]{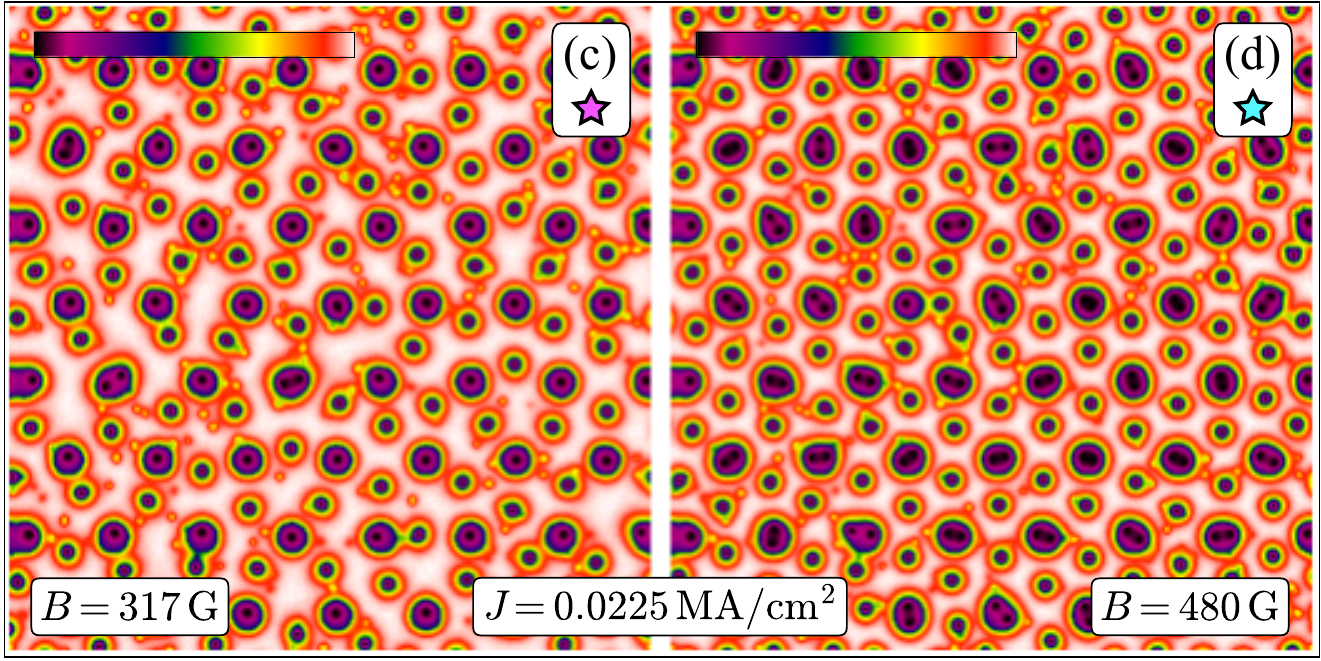}}\label{fig:psi_h12nm_B317G_J0p0225MAcm2}}
	\subfloat{\label{fig:psi_h12nm_B480G_J0p0225MAcm2}}
	\end{center} \vspace{-4mm}
	\caption{
		\protect\subref{fig:psi_h12nm_B108G_J0p108MAcm2}--\protect\subref{fig:psi_h12nm_132G_J0p108MAcm2}.~Snapshots 
		of the \uhref{https://youtu.be/OOzHO7OTqSg}{vortex dynamics} for applied 
		current $J = 0.108$\,MA/cm$^2$ and magnetic fields $B = 108$\,G 
		(\protect\subref{fig:psi_h12nm_B108G_J0p108MAcm2}) 
		and $132\,$G (\protect\subref{fig:psi_h12nm_132G_J0p108MAcm2}) 
		marked by yellow and blue stars in Fig.~\ref{fig:Jc_B_exp_sim}. 
		The pink color corresponds to the superconducting state, while black 
		indicates the complete suppression of the superconducting order parameter. 
		Since~$B$ is close to $\Bm$, almost every pinning center contains a single vortex. 
		The added background weak pinning centers appear as small yellow dots. 
		In panel~\protect\subref{fig:psi_h12nm_B108G_J0p108MAcm2}, four inclusions 
		are not occupied by vortices (indicated by green arrows) and one vortex 
		is stacked in the center of triangular cell (black arrow). 
		In panel~\protect\subref{fig:psi_h12nm_132G_J0p108MAcm2}, 
		five vortices are not pinned and drift.
		\protect\subref{fig:psi_h12nm_B317G_J0p0225MAcm2}--\protect\subref{fig:psi_h12nm_B480G_J0p0225MAcm2}.~\uhref{https://youtu.be/TAfHOG_lfEk}{Vortex dynamics} 
		for magnetic field $B = 317$\,G (\protect\subref{fig:psi_h12nm_B317G_J0p0225MAcm2}) 
		and $480\,$G (\protect\subref{fig:psi_h12nm_B480G_J0p0225MAcm2}) 
		at applied current $J = 0.0225$\,MA/cm$^2$ for positions marked 
		by the magenta and cyan stars in Fig.~\ref{fig:Jc_B_exp_sim}.
		In panel~\protect\subref{fig:psi_h12nm_B317G_J0p0225MAcm2}, 
		most of the inclusions contain two vortices and vortices outside 
		the inclusions form a honeycomblike lattice.
	}
	\label{fig:psi_sameJ}
\end{figure}

Next, we compare the vortex dynamics near the drop at $B = \Bm$. Snapshots of the order parameter amplitude $|\psi|^2$ are shown in Fig.~\ref{fig:psi_h12nm_B108G_J0p108MAcm2} for $B = 0.9\Bm = 108$\,G and Fig.~\ref{fig:psi_h12nm_132G_J0p108MAcm2} for $B = 1.1\Bm = 132$\,G; these values are marked by yellow and blue stars in Fig.~\ref{fig:Jc_B_exp_sim}, accordingly. (The corresponding vortex dynamics can be seen in the supplementary \uhref{https://youtu.be/OOzHO7OTqSg}{video}.) The external current is the same for both snapshots and chosen to be $J = 0.108\,\mathrm{MA}/\mathrm{cm}^2 \sim [\Jc(0.9\Bm) \Jc(1.1\Bm)]^{1/2}$. In these density plots, which are $xy$ cross sections at fixed $z = 17$\,nm, an amplitude of $|\psi|^2 \sim 1$ or ``full'' superconductivity corresponds to white regions, while the order parameter is completely suppressed in black areas with $|\psi|^2 \sim 0$. Since the chosen cross section is slightly underneath the inclusion cylinder pieces, they appear violet with $|\psi|^2 \sim 0.1$ due to the proximity effect. Since the order parameter is completely suppressed in the vortex cores, they appear black; thus one can distinguish occupied and unoccupied inclusions. Figure~\ref{fig:psi_h12nm_B108G_J0p108MAcm2} with $B < \Bm$ and $J < \Jc$ shows four unoccupied inclusions and one ``caged'' vortex pinned by repulsive forces at the center of the triangular cell. In Fig.~\ref{fig:psi_h12nm_B108G_J0p108MAcm2} with $B > \Bm$ and $J > \Jc$, all of the inclusions pin one vortex and five nearly free vortices drift between the inclusions and produce a finite voltage across the sample.

\begin{figure}
	\begin{center}
	\subfloat{\href{https://youtu.be/iLfV3d5BZug}{\includegraphics[width=8.7cm]{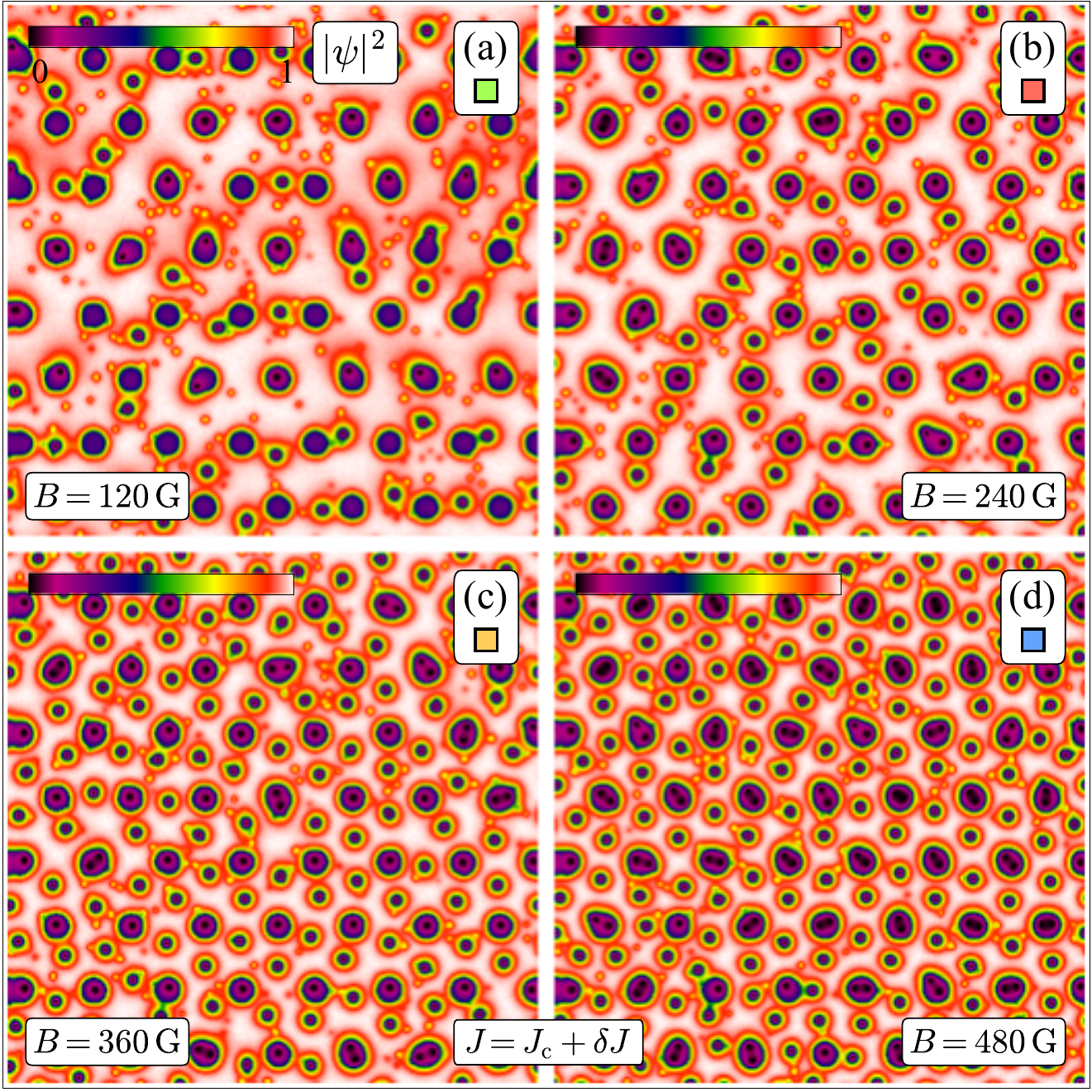}}\label{fig:psi_h12nm_B120G_dJ}}
	\subfloat{\label{fig:psi_h12nm_B240G_dJ}}
	\subfloat{\label{fig:psi_h12nm_B360G_dJ}}
	\subfloat{\label{fig:psi_h12nm_B480G_dJ}}
	\end{center} \vspace{-4mm}
	\caption{
		Snapshots of the \uhref{https://youtu.be/iLfV3d5BZug}{order parameter dynamics} 
		for magnetic fields at the matching fields 
		$B = n\Bm = 120$, $240$, $360$, $480$\,G 
		(with $n = 1$ to $4$ for panels \protect\subref{fig:psi_h12nm_B120G_dJ} 
		to \protect\subref{fig:psi_h12nm_B480G_dJ}, respectively) and the applied 
		current slightly larger than the critical current $J$ at given field, 
		$J = \Jc(B) + \delta J$. 
		We chose $\delta J$ as small as possible and adjusted the frame rate to 
		have approximately the same vortex drifting speed of vortices in all panels.
		These values of fields and current are marked by colored squares 
		in Fig.~\ref{fig:Jc_B_exp_sim}.
		In panel~\protect\subref{fig:psi_h12nm_B120G_dJ}, each inclusion pins 
		a single vortex and, after depinning, the vortex lattice moves as a whole.
		In panel~\protect\subref{fig:psi_h12nm_B240G_dJ}, half of the triangular 
		cells making up the pinscape contain one vortex.
		During the depinning process, these ``free'' vortices individually move 
		to neighboring unoccupied cells in the pinscape. This process results 
		in the suppression of the critical current by $\sim 3.5$ times compared 
		to the first matching filed. For the third and fourth matching fields shown 
		in panels \protect\subref{fig:psi_h12nm_B360G_dJ} and 
		\protect\subref{fig:psi_h12nm_B480G_dJ}, vortices located between 
		inclusions form an almost perfect honeycomb lattice. 
		Additionally, in the situation depicted 
		in panel~\protect\subref{fig:psi_h12nm_B480G_dJ}, each pinning 
		center nominally pins two vortices, thereby increasing the stiffness 
		of the caged vortices, resulting in a higher critical current compared 
		to panel~\protect\subref{fig:psi_h12nm_B360G_dJ}. 
	}
	\label{fig:psi_dJ}
\end{figure}

\begin{figure}
	\begin{center}
	\subfloat{\href{https://youtu.be/IWGVXDG86-w}{\includegraphics[width=8.7cm]{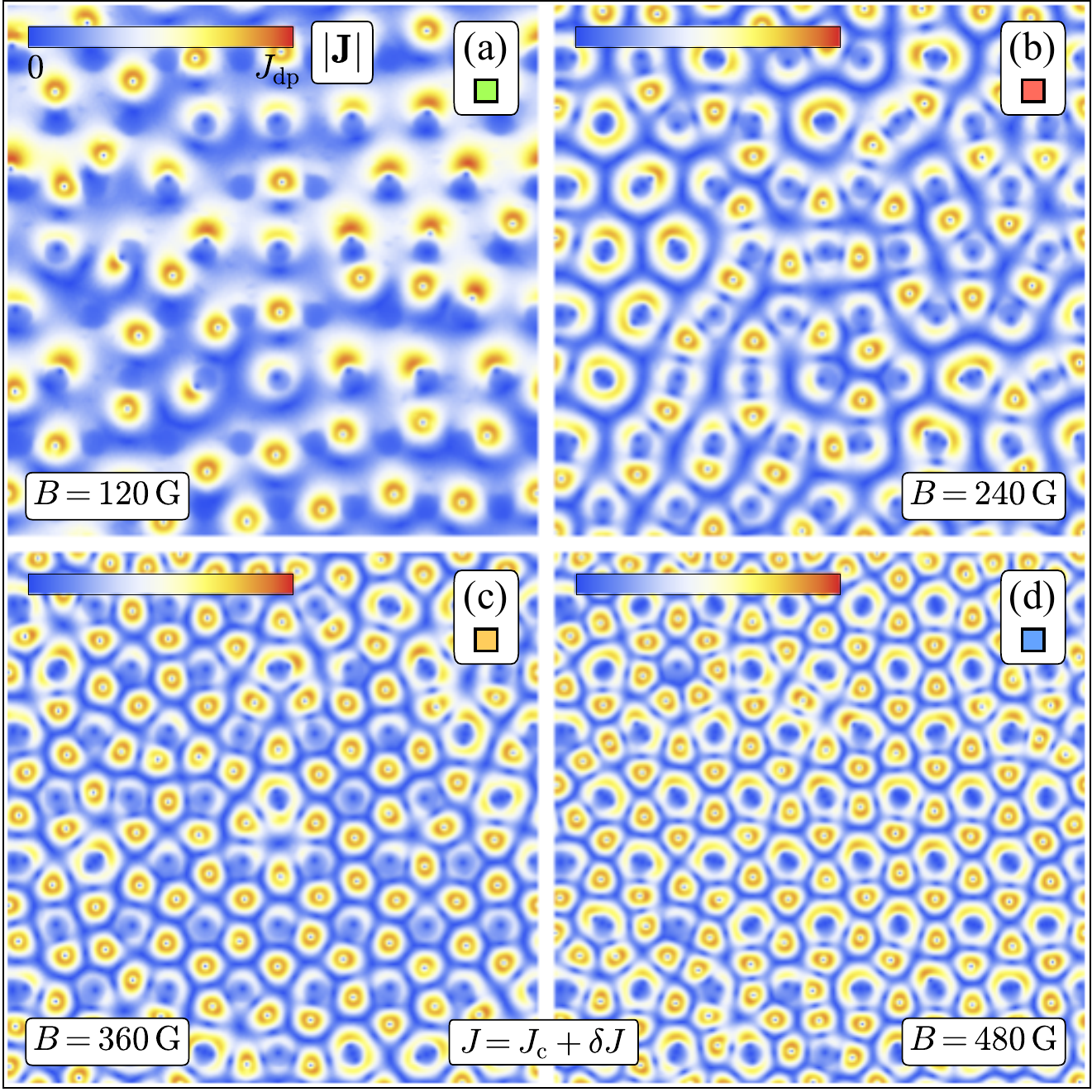}}\label{fig:J_h12nm_B120G_dJ}}
	\subfloat{\label{fig:J_h12nm_B240G_dJ}}
	\subfloat{\label{fig:J_h12nm_B360G_dJ}}
	\subfloat{\label{fig:J_h12nm_B480G_dJ}}
	\end{center} \vspace{-4mm}
	\caption{
		Snapshots of the \uhref{https://youtu.be/IWGVXDG86-w}{supercurrent dynamics} 
		for the same parameters as in Fig.~\ref{fig:psi_dJ}.
	}
	\label{fig:J_dJ}
\end{figure}

With increasing magnetic field, the number of vortices in between and on the pinning site increases. At a fixed applied current, we compare the order parameter amplitude for two values of the magnetic field, $B = 2.64\Bm = 317$\,G and $B = 4\Bm = 480$\,G, shown in Figs.~\ref{fig:psi_h12nm_B317G_J0p0225MAcm2} and \ref{fig:psi_h12nm_B480G_J0p0225MAcm2}, respectively. (See also the supplementary \uhref{https://youtu.be/TAfHOG_lfEk}{video} showing the vortex dynamics.) The external current is $J = 0.0225\,\mathrm{MA}/\mathrm{cm}^2$. The corresponding current and magnetic field values are marked by magenta and cyan crosses in the $J$-$B$ diagram; see Fig.~\ref{fig:Jc_B_exp_sim}. For these two parameter sets, the vortex matter behavior is quite different from Figs.~\ref{fig:psi_h12nm_B108G_J0p108MAcm2} and \ref{fig:psi_h12nm_132G_J0p108MAcm2}. Indeed, in Fig.~\ref{fig:psi_h12nm_B317G_J0p0225MAcm2} with $J > \Jc$, inclusions pin one or two vortices, but ``caged'' vortices can ``squeeze'' between the inclusions generating dissipation. In Fig.~\ref{fig:psi_h12nm_B480G_J0p0225MAcm2} with $J < \Jc$, almost all inclusions pin two vortices each and caged vortices form a honeycomb lattice. 

In this regime, the repulsion between vortices at inclusions and caged vortices is so strong that the latter cannot squeeze through the hexagonal pattern of inclusions and, consequently, results in the peak at $B = 4\Bm$. The vortex lattice structure at this field, corresponding to four times the matching field, is defined by two vortices per inclusion ($2\Bm$) and a hexagonal lattice in between them ($2\Bm$). This is a typical scenario: if one can associate two caged vortices to one inclusion and the inclusion pins $n - 2$ vortices, one observes a peak at the $n$-th matching field, $B = n\Bm$. For $n = 3$, each inclusion pins one vortex and cages two (see, e.g., blue curve in Fig.~\ref{fig:Jc_B_Tci_h12nm} below); for $n = 4$ each inclusion pins two vortices (red curve in Fig.~\ref{fig:Jc_B_exp_sim}); and for $n = 5$, three vortices are pinned by an inclusion (blue curve in Fig.~\ref{fig:Jc_B_h} below). Peaks appearing at $n = 1$ or $n = 2$ can be related to single or double occupation of pinning centers. 

Before the $n$-th peak ($n > 2$), at the preceding $(n - 1)$-th matching magnetic field, $B = (n - 1)\Bm$, the critical current has a smaller value, $\Jc([n - 1]\Bm) < \Jc(n\Bm)$. In this regime, one caged vortex per inclusion is missing. These vacancies can be easily occupied by neighboring caged vortices and therefore allow vortices to move one by one by the Lorentz force. This mechanism significantly reduces the critical current at the $(n - 1)$-th matching magnetic field and is more pronounced for stronger pinning centers. 

\begin{figure}
	\begin{center}
		\subfloat{\hspace{-0.25cm} \includegraphics[width=8.9cm]{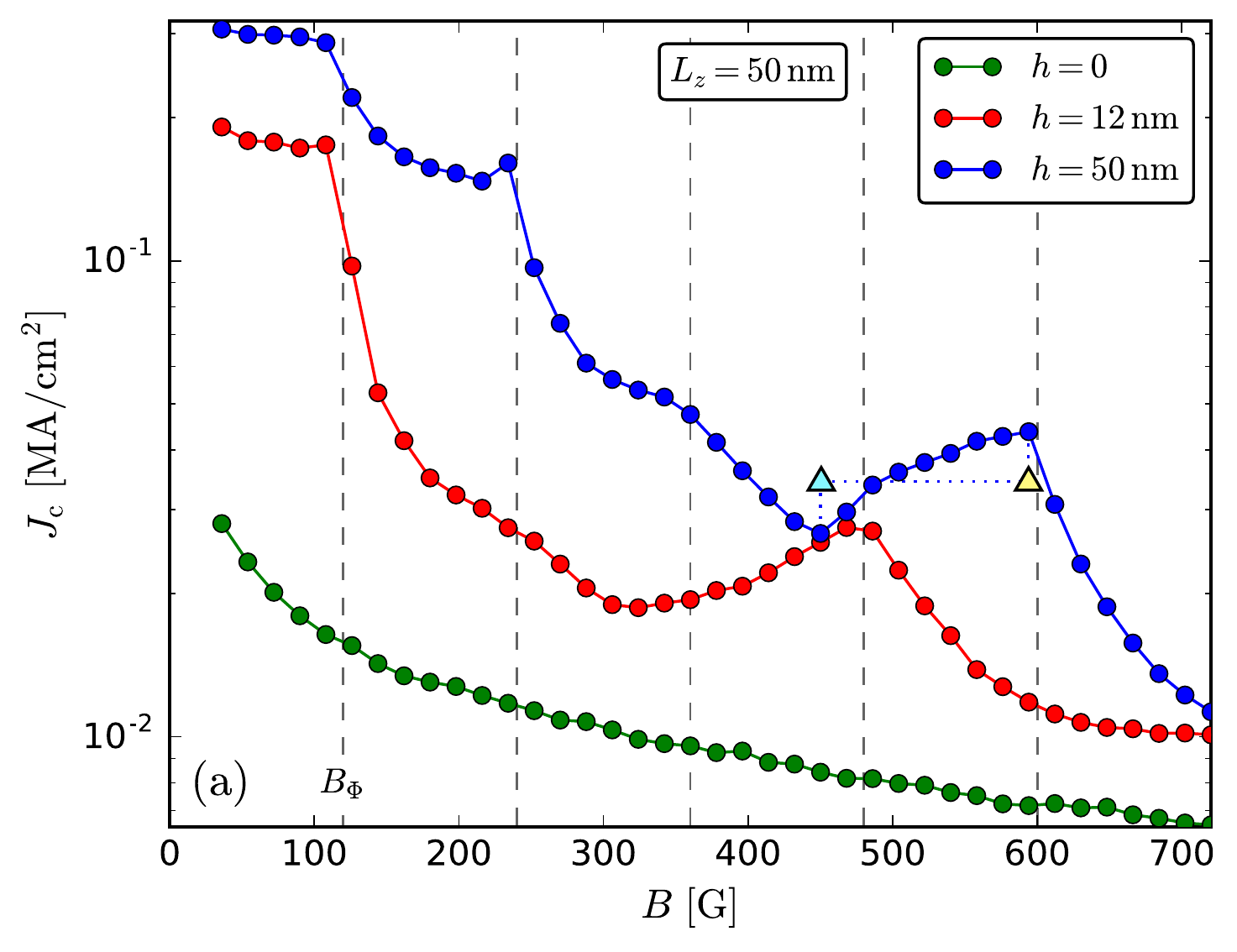} \label{fig:Jc_B_h}} \vspace{-4mm}
		\subfloat{\hspace{-0.15cm} \href{https://youtu.be/CIez7dOzkmg}{\includegraphics[width=8.7cm]{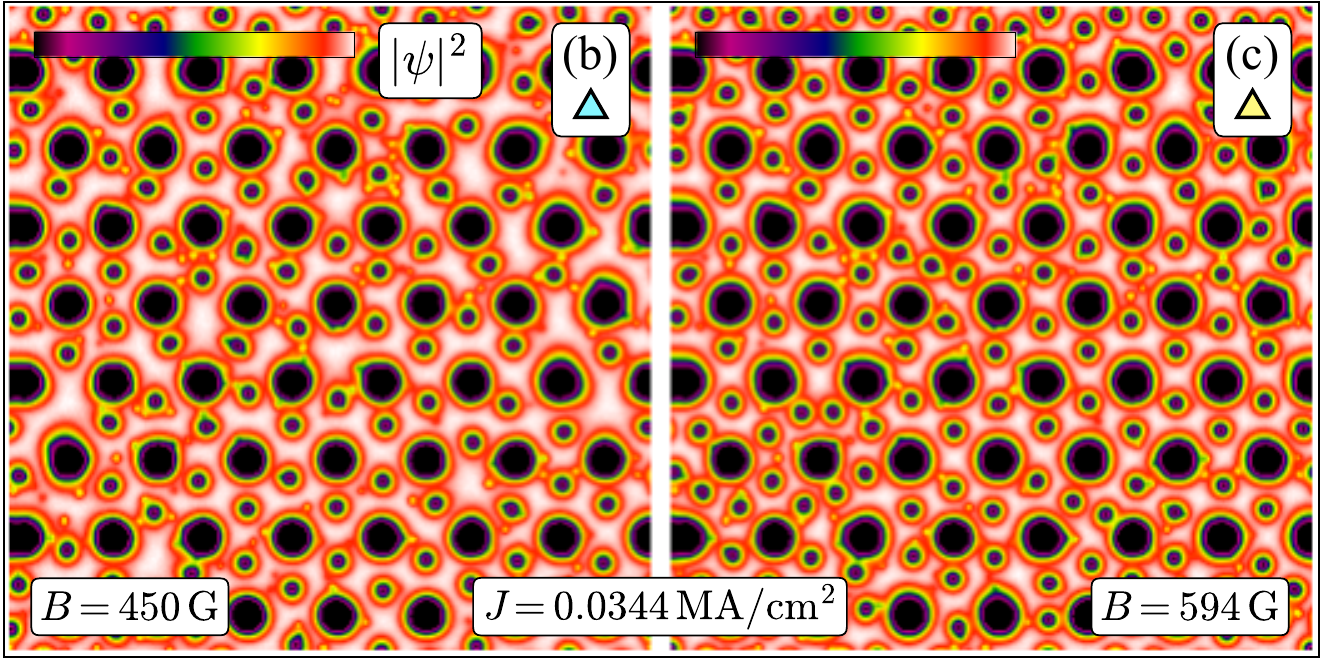}} \label{fig:psi_h50nm_B450G_J0p0344MAcm2}}
		\subfloat{\label{fig:psi_h50nm_B594G_J0p0344MAcm2}}
	\end{center} \vspace{-5mm}
	\caption{ 
		\protect\subref{fig:Jc_B_h}.~Numerical simulation of the $\Jc(B)$ dependence 
		on inclusion depth~$h$ and $\Tci = 0$. The strongest possible inclusions (blue) 
		are drilled through-holes $h=50$\,nm with zero critical temperature. 
		They generate peaks at $B = 2\Bm$ and $5\Bm$. 
		The simulated $\Jc(B)$ curve has overall higher values compared 
		to experimental samples with $h=12$\,nm (blue) and the peak 
		at $4\Bm$ disappears.
		\protect\subref{fig:psi_h50nm_B450G_J0p0344MAcm2}--\protect\subref{fig:psi_h50nm_B594G_J0p0344MAcm2}.~Snapshots 
		of the \uhref{https://youtu.be/CIez7dOzkmg}{vortex dynamics} 
		for an inclusion depth of $h = 50$\,nm, applied current 
		$J = 0.0344$\,MA/cm$^2$, and magnetic fields $B = 450$\,G 
		(\protect\subref{fig:psi_h50nm_B450G_J0p0344MAcm2}) and 
		$594\,$G (\protect\subref{fig:psi_h50nm_B594G_J0p0344MAcm2}) 
		marked by cyan and yellow triangles in panel~\protect\subref{fig:Jc_B_h}.
	}
\end{figure}

Figures~\ref{fig:psi_dJ} and \ref{fig:J_dJ} as well as the supplementary videos show snapshots of squared order parameter $|\psi|^2$ configurations and supercurrent for external current slightly above the critical current near the first four matching fields, $J = \Jc + \delta J$ with $0 < \delta J \ll \Jc$,  illustrating the different pinning behavior and dynamics. (See \uhref{https://youtu.be/iLfV3d5BZug}{video of dynamics of squared order parameter amplitude} and \uhref{https://youtu.be/IWGVXDG86-w}{supercurrent}, \uhref{https://youtu.be/GH1vsZqIzXw}{combination of those}, and \uhref{https://youtu.be/xa7x8yS53SA}{combination with order parameter phase}.) At the first matching field (120\,G, one vortex per inclusion), vortices move mostly by jumping from one inclusion to another; see Fig.~\ref{fig:psi_h12nm_B120G_dJ}. In contrast to that, near the second matching field, some doubly occupied defects lose one vortex, which then moves in between the defects; see Fig.~\ref{fig:psi_h12nm_B240G_dJ}. At the two higher fields, Figs.~\ref{fig:psi_h12nm_B360G_dJ} and \ref{fig:psi_h12nm_B480G_dJ}, vortices are also pinned in between the defects due to their repulsive interaction and defects have, at most, two pinned vortices in the stationary state. Above the third matching field, the vortex dynamics is still characterized by meandering motion in between the defects, while at the fourth matching field, they move via absorption and emission at the defects, which requires larger currents than for the second and third matching field. The higher value of the critical current at the fourth matching field is therefore a result of a ``blocking'' behavior of the vortices, which are pinned between inclusions.

\subsection{Dependence on inclusion strength} \label{sec:results_strength}

Above we discussed a particular realization of the pinning array, which, at least qualitatively, explains the experimentally observed field dependence of the critical current. However, it is instructive to study the dependence of the overall behavior on the inclusion strength. In the framework of the chosen model, we can control the strength of the inclusions by changing the depth $h$ of the inclusions, the critical temperature $\Tci$ inside the inclusions, and their diameter $d$. All three quantities influence the pinning strength of the inclusions and one can expect that the effects from increasing $h$ are similar to the effects from decreasing $\Tci$.

\paragraph*{Inclusion depth.}
The $\Jc(B)$ curve for different inclusion depths $h$ of inclusions is shown in Fig.~\ref{fig:Jc_B_h}. The green and red curves correspond to $h = 0$ (reference) and $h = 12$\,nm (25\% of the $50$\,nm sample thickness). The same curves were compared to experiment in Fig.~\ref{fig:Jc_B_exp_sim}. Surprisingly, the positions of the peaks in $\Jc(B)$ change with $h$. Indeed, the blue curve in Fig.~\ref{fig:Jc_B_h} for $h = 50$\,nm demonstrates two prominent peaks at $B = 2\Bm$ and $5\Bm$ and has a local minimum near $4\Bm$.

The difference between fully and partially drilled holes in the sample can be analyzed in the same way as in Sec.~\ref{sec:results_comparison}. Snapshots of $|\psi|^2$ density plots are shown in Fig.~\ref{fig:psi_h50nm_B450G_J0p0344MAcm2} for $B = 3.75\Bm = 450$\,G and Fig.~\ref{fig:psi_h50nm_B594G_J0p0344MAcm2} for $B = 4.95\Bm = 594$\,G (magenta and cyan crosses in Fig.~\ref{fig:Jc_B_h}); see also supplementary \uhref{https://youtu.be/CIez7dOzkmg}{video} of vortex dynamics. The external current is fixed at $J = 0.0344\,\mathrm{MA}/\mathrm{cm}^2$. The vortex dynamics is very similar to the one depicted in Figs.~\ref{fig:psi_h12nm_B317G_J0p0225MAcm2} and \ref{fig:psi_h12nm_B480G_J0p0225MAcm2}. The only difference is that in Fig.~\ref{fig:psi_h50nm_B450G_J0p0344MAcm2}, each inclusion is stronger and pins three vortices. Therefore, the resulting vortex configuration gives rise to a peak at five times the matching field, having three vortices per inclusion and two caged vortices (as for $h = 12$\,nm) per pattern unit cell.

\begin{figure}
	\begin{center}
		\subfloat{\hspace{-0.25cm} \includegraphics[width=8.9cm]{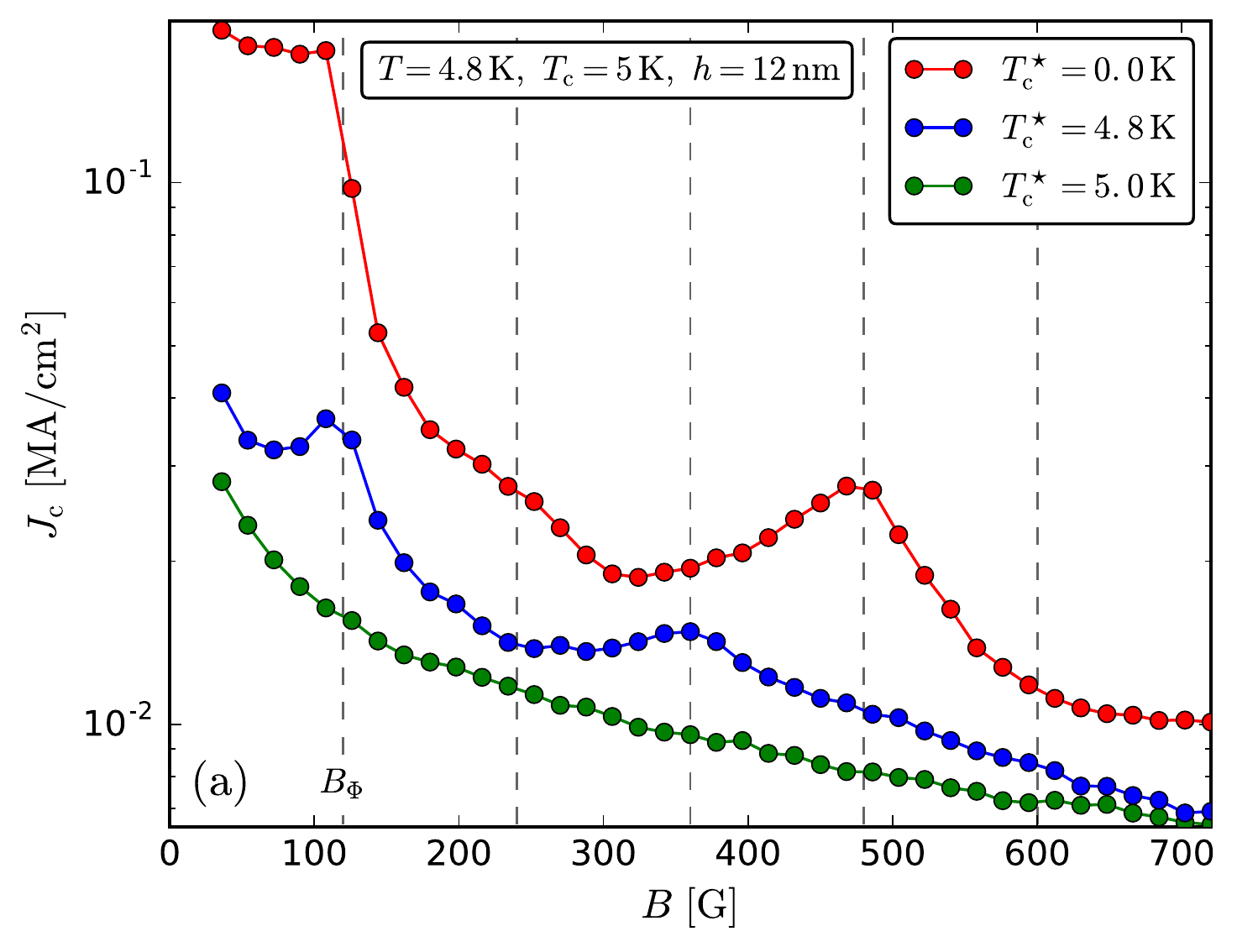} \label{fig:Jc_B_Tci_h12nm}} \vspace{-4mm} \newline
		\subfloat{\hspace{-0.25cm} \includegraphics[width=8.9cm]{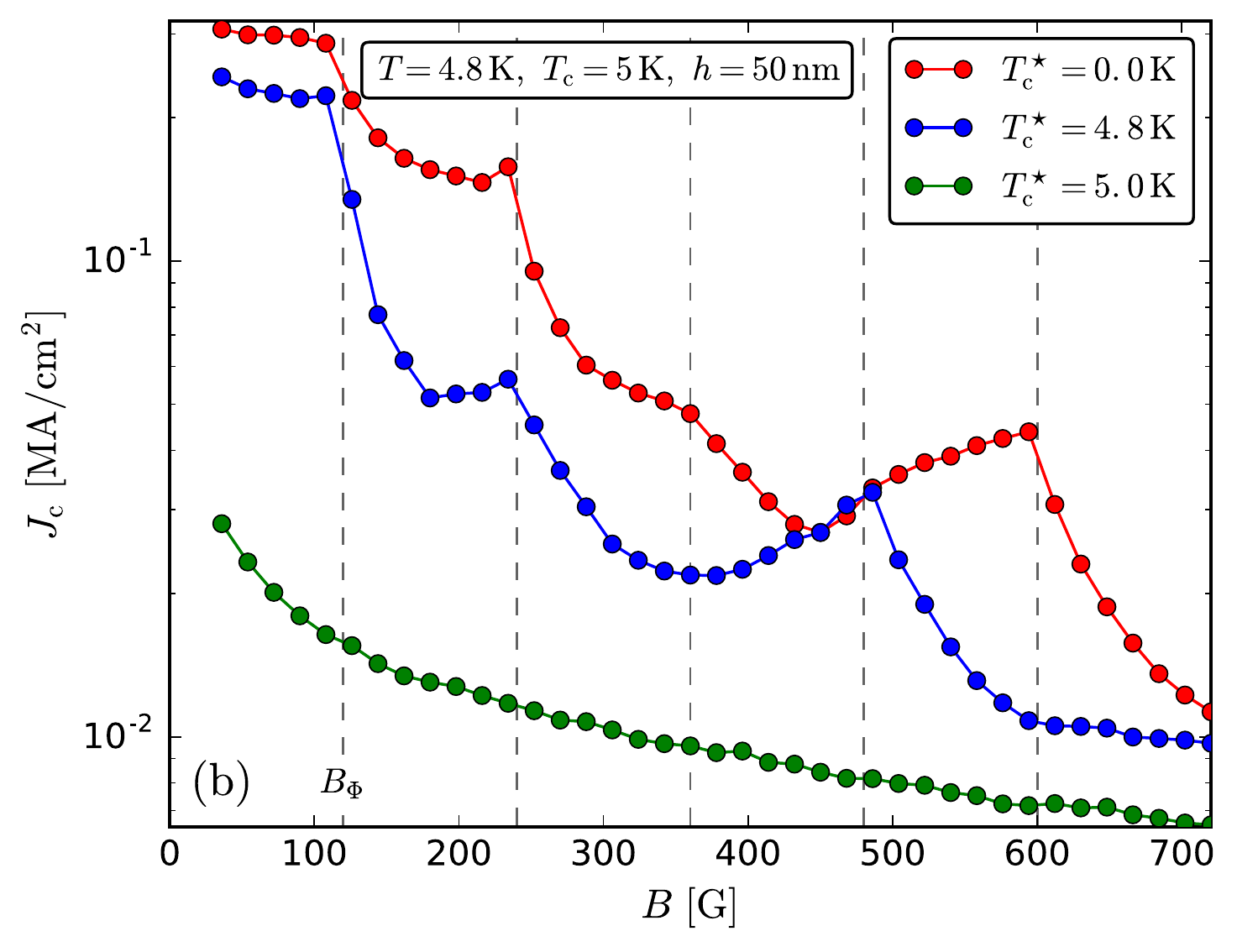} \label{fig:Jc_B_Tci_h50nm}} \vspace{-2mm}
	\end{center} \vspace{-5mm}
	\caption{
		\protect\subref{fig:Jc_B_Tci_h12nm}.~Numerical $\Jc(B)$ dependence 
		for different critical temperatures inside inclusions $\Tci$ with inclusion 
		depth $h = 12$\,nm. 
		\protect\subref{fig:Jc_B_Tci_h50nm}.~The same for through-hole 
		inclusions, $h = 50$\,nm.
		In the case of strongest pinning $\Tci = 0$, one observes peaks 
		in $\Jc(B)$ at $B = 2\Bm$ and $5\Bm$. However, for different $\Tci$, 
		the positions of the peaks change. For example, at 
		$\Tci = 4.8\,\mathrm{K} = T$ one can see two peaks at $2\Bm$ and 
		$4\Bm$.
	}
\end{figure}

\paragraph*{Critical temperature inside inclusion.}
The dependence of the critical temperature $\Tci$ inside the inclusion of depth $h = 12$\,nm is shown in Fig.~\ref{fig:Jc_B_Tci_h12nm}. Naturally, increasing $\Tci$ reduces the pinning strength and therefore the critical current (compare the red curve for $\Tci = 0$ and blue curve for $\Tci = 4.8$\,K). $\Jc(B)$ curves for inclusions penetrating the entire film, i.e., $h = 50$\,nm, are shown in Fig.~\ref{fig:Jc_B_Tci_h50nm}. As before, we see that increasing the pinning depth shifts the peak from $B = 4\Bm$ (red curve in Fig.~\ref{fig:Jc_B_Tci_h12nm}) to $5\Bm$ (red curve in Fig.~\ref{fig:Jc_B_Tci_h50nm}). However, increasing the critical temperature inside the inclusions from $\Tci = 0$ to $4.8$\,K moves the peak back to $B = 4\Bm$ due to decreasing pinning strength of the inclusions (blue curve in Fig.~\ref{fig:Jc_B_Tci_h50nm}). 

\begin{figure}
	\begin{center}
		\subfloat{\hspace{-0.25cm} \includegraphics[width=8.9cm]{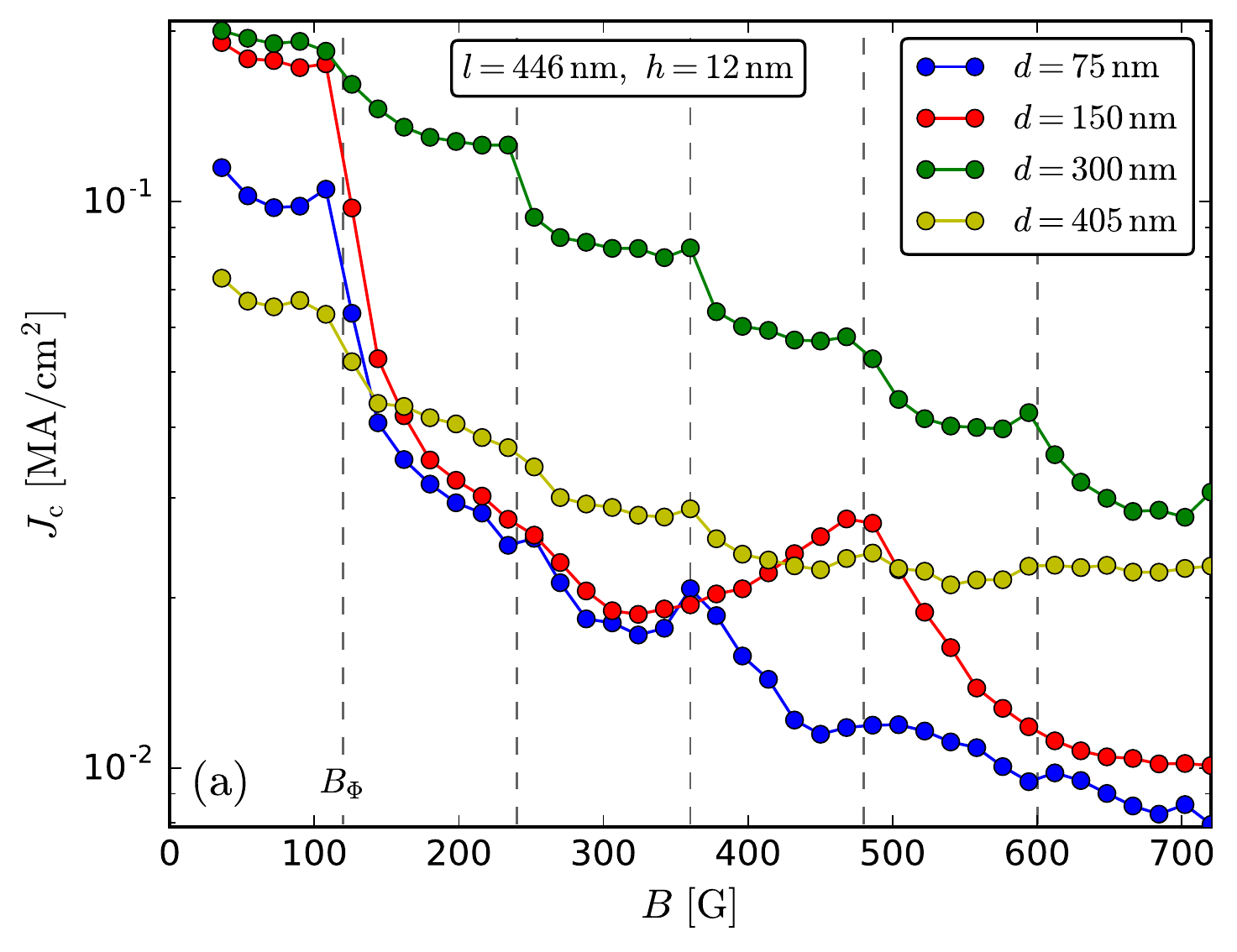} \label{fig:Jc_B_d_h12nm}} \vspace{-2mm} \newline
		\subfloat{\includegraphics[width=8.6cm]{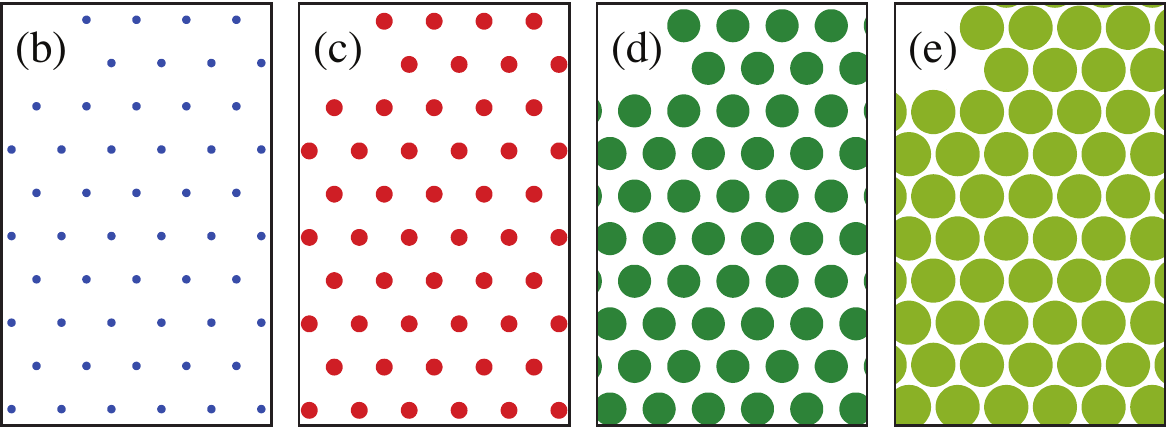} \label{fig:hexagonal_d3}} \newline
		\subfloat{\label{fig:hexagonal_d6}}
		\subfloat{\label{fig:hexagonal_d12}}
		\subfloat{\label{fig:hexagonal_d16p2}} \newline
	\end{center} \vspace{-5mm}
	\caption{ 
		\protect\subref{fig:Jc_B_d_h12nm}.~$\Jc(B)$ dependence 
		for different inclusion diameter $d$.
		The color-coded corresponding patterns are shown in 
		panel~\protect\subref{fig:hexagonal_d3} for $d = 75$\,nm,
		panel~\protect\subref{fig:hexagonal_d6} for $d = 150$\,nm,
		panel~\protect\subref{fig:hexagonal_d12} for $d = 300$\,nm, and
		panel~\protect\subref{fig:hexagonal_d16p2} for $d = 405$\,nm.
	}
\end{figure}

\paragraph*{Dependence on inclusion diameter.}
Figure~\ref{fig:Jc_B_d_h12nm} shows the $\Jc(B)$ dependences for different inclusion diameters. The corresponding hexagonal patterns are displayed in Figs.~\ref{fig:hexagonal_d3}--\ref{fig:hexagonal_d16p2} by the same colors. As one can see, the strength of the inclusions increases with their diameter until their volume fraction occupies too much of the sample. Indeed, pinning centers with relatively small diameter, $d = 3\xi = 75$\,nm, generate low critical current, as shown by the blue curve in Fig.~\ref{fig:Jc_B_d_h12nm}. These inclusions may pin one vortex only; thus we observe a peak at $B = 3\Bm$ ($\Bm$ from pinned vortices and $2\Bm$ from caged vortices). For the increased diameter, $d = 6\xi = 150$\,nm, the peak shifts towards $4\Bm$ (red curve) as was discussed above. 

The larger diameter, i.e., $d = 12\xi = 300$\,nm, creates a stronger pinning force resulting in higher $\Jc$ (green curve in Fig.~\ref{fig:Jc_B_d_h12nm}). Concurrently, the larger pinning center may accommodate one to approximately eight vortices, depending on the applied magnetic field. This behavior is reflected by the series of peaks at each integer multiple of~$\Bm$. In this regime, the system shows behavior similar to Josephson junction arrays frustrated by magnetic field. However, further increasing the inclusion diameter reduces~$\Jc$ (see yellow curve for $d = 405$\,nm). The main reason for the decrease in $\Jc$ is the reduced superconductivity in the gaps between the inclusions, $l - d = 1.6\xi = 41$\,nm, which in turn reduces the effective potential barrier for hopping of vortices between inclusions. Additionally, large inclusion diameters simply decrease the effective superconductor cross section for supercurrent transport. One sees that the matching effects become much weaker in this case.

To summarize, we conclude that the vortex dynamics responsible for the critical current depends on two main factors: (i) matching field $\Bm$ or hexagonal pattern constant $l$ and (ii) pinning strength of the inclusion. The latter depends on inclusion depth $h$, critical temperature inside inclusion $\Tci$, and inclusion diameter $d$. These three quantities are responsible for the pinning force of the inclusion and its ability to pin more than one vortex.

\subsection{Influence of randomness} \label{sec:results_randomness}

\begin{figure}
	\begin{center}
		\subfloat{\hspace{-0.25cm} \includegraphics[width=8.9cm]{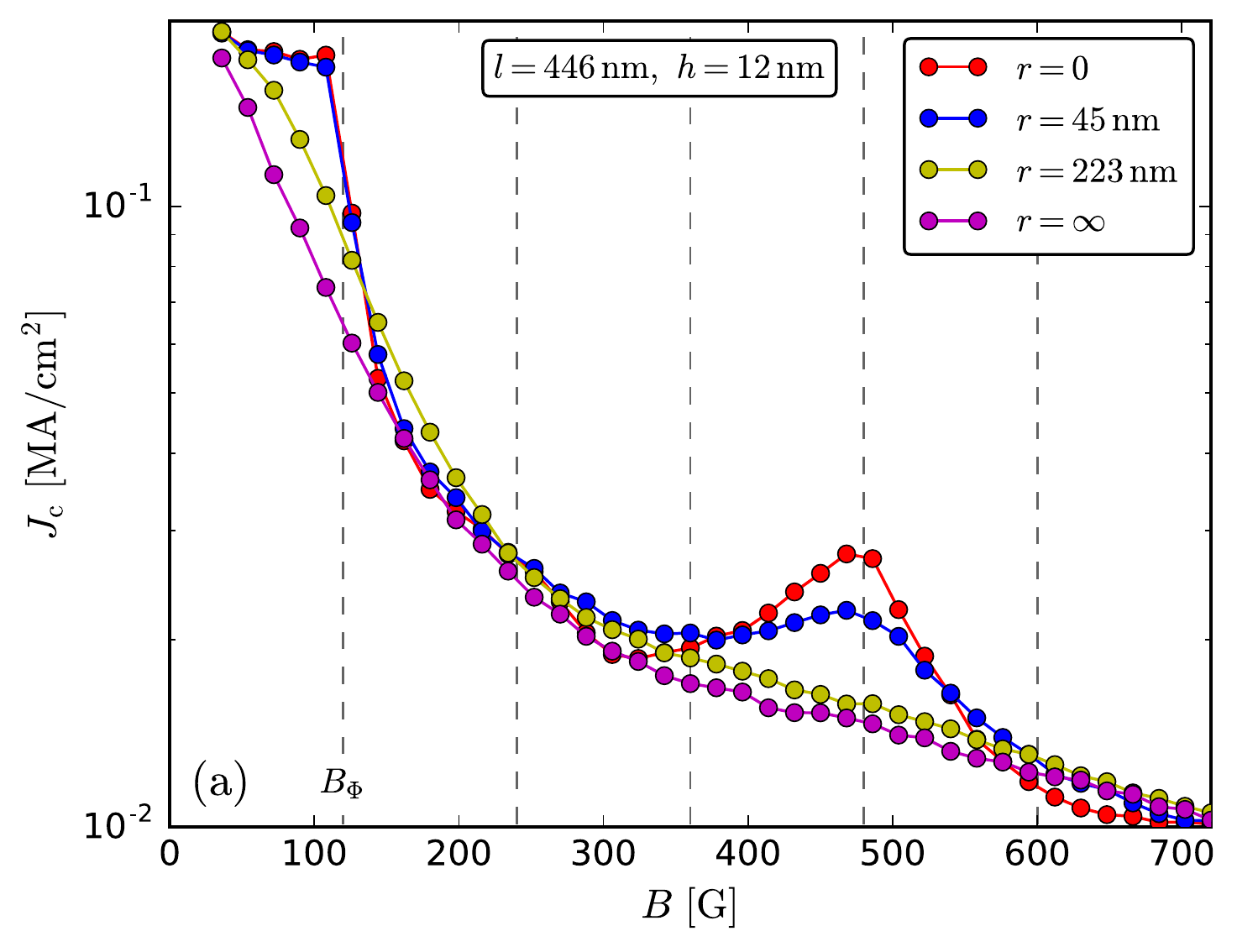} \label{fig:Jc_B_r_h12nm}} \vspace{-2mm} \newline
		\subfloat{\hspace{-0.25cm} \includegraphics[width=8.9cm]{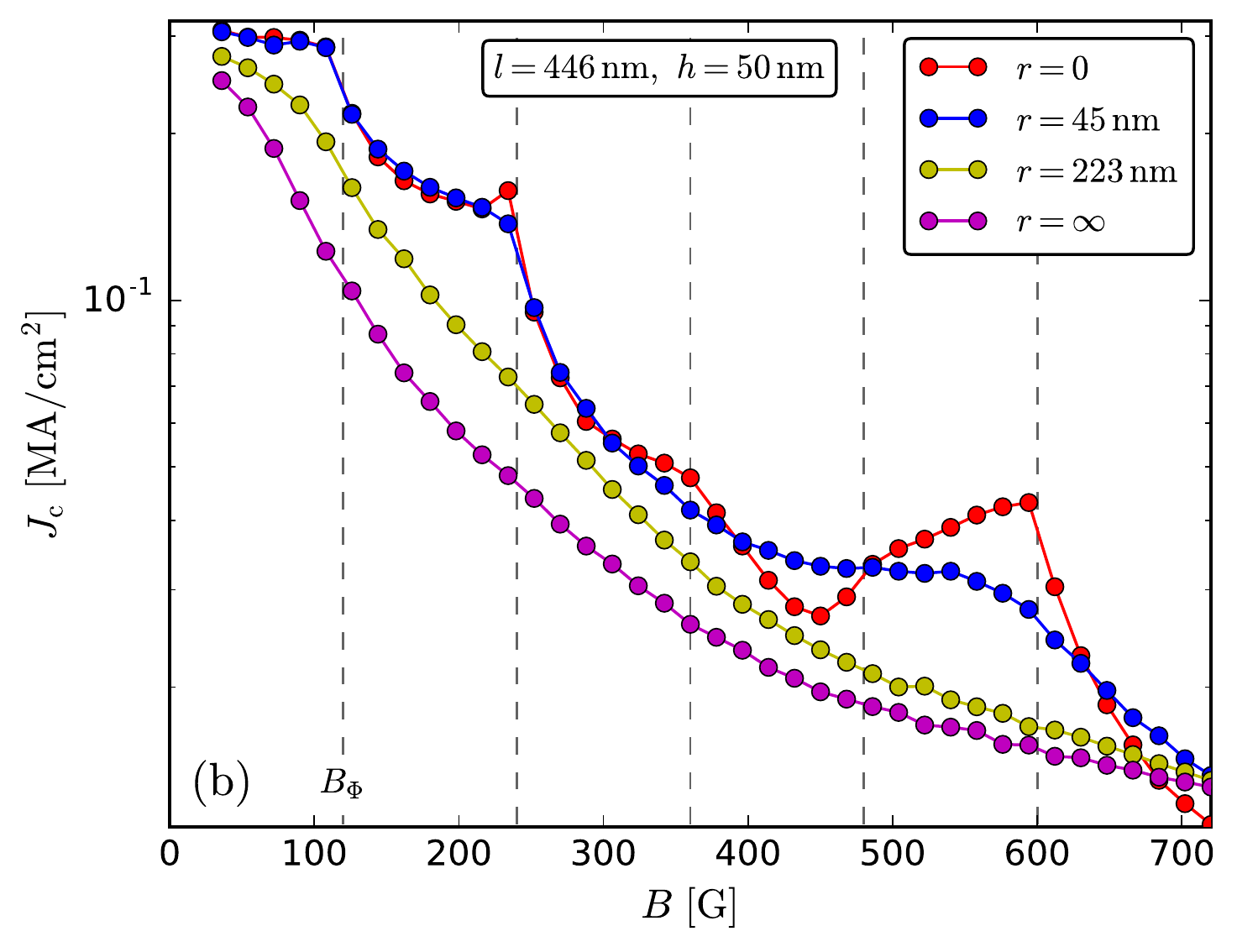} \label{fig:Jc_B_r_h50nm}} \vspace{-2mm} \newline
		\subfloat{\includegraphics[width=8.6cm]{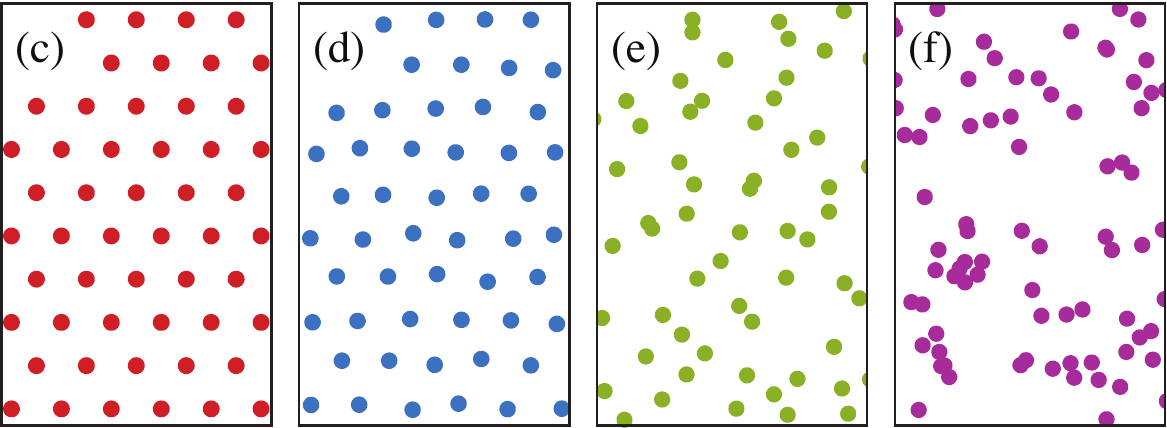} \label{fig:hexagonal_r0}}
		\subfloat{\label{fig:hexagonal_r45}}
		\subfloat{\label{fig:hexagonal_r223}}
		\subfloat{\label{fig:hexagonal_rinf}}
	\end{center} \vspace{-5mm}
	\caption{ 
		\protect\subref{fig:Jc_B_r_h12nm}.~Numerical $\Jc(B)$ dependence 
		with added randomness $r$ to the inclusion positions, compared 
		with an ideal hexagonal lattice with lattice constant $l = 446\,$nm and inclusion 
		depth $h = 12$\,nm. For ideal ($r = 0$, red) and slightly perturbed 
		($r = 75\,$nm, blue) lattice, one sees the drop 
		at $B = \Bm$ and the peak at $4\Bm$. 
		For a randomness value of $r = 188\,$nm, which is comparable 
		with the lattice constant (case of hyperuniform order), the 
		$\Jc(B)$ dependence becomes monotonic (yellow). 
		The uncorrelated inclusion position distribution ($r$ larger than the system size) 
		leads to a monotonic $\Jc(B)$ dependence (magenta), which 
		is smaller compared with the previous case.
		\protect\subref{fig:Jc_B_r_h50nm}.~For through-hole inclusions, $h = 50$\,nm, 
		the difference in critical current between hyperuniform (yellow) and 
		uncorrelated (magenta) inclusion placements becomes more pronounced.		
		\protect\subref{fig:hexagonal_r0}.~Perfect hexagonal pattern.
		\protect\subref{fig:hexagonal_r45}.~Weak randomness.
		\protect\subref{fig:hexagonal_r223}.~Hyperuniform order characterized 
		by randomness at short scales and homogeneity at long scales.
		\protect\subref{fig:hexagonal_rinf}.~Uncorrelated random placement.
	}
\end{figure}

Finally, we discuss the influence of random displacements of the inclusions from their perfect lattice positions. This gives an insight on when matching effects become relevant and shows how possible lattice imperfections in experimental systems affect the vortex dynamics. Let us start with the perfect hexagonal lattice (Fig.~\ref{fig:hexagonal_r0}) and add some random displacement within an interval $[-r \ldots r]$ to the $x$ and $y$ positions of each inclusion. The resulting patterns are presented in Figs.~\ref{fig:hexagonal_r45}--\ref{fig:hexagonal_rinf}. $\Jc(B)$ curves are shown in Fig.~\ref{fig:Jc_B_r_h12nm} for $h = 12$\,nm and in Fig.~\ref{fig:Jc_B_r_h50nm} for $h = 50$\,nm in corresponding colors.

The perfect hexagonal pattern (red lines in Figs.~\ref{fig:Jc_B_r_h12nm} and \ref{fig:Jc_B_r_h50nm}), i.e., pattern with well-defined matching field~$\Bm$, demonstrate the most nonmonotonic behavior with peaks at some integer multiples of $\Bm$ as discussed before. With increasing randomness in the placement of the inclusions, $r$, these peaks become less pronounced (blue). For some random placement, peaks or kinks near integer multiples of $\Bm$ completely disappear (yellow) and, as a result, the $\Jc(B)$ curves become more monotonic. The values of the critical current for moderate randomness may be both lower or higher compared to the ordered pattern. 

This order of defects is sometimes called \textit{hyperuniformity} and is characterized by uniform defect placement on large scales and disordered placement on small scales. This hyperuniform order is characterized by higher critical current compared to uncorrelated defect placement (magenta). This result is in agreement with recent Langevin-dynamics analysis of pinning.\cite{Thien:2016} In terms of critical current, there are two main differences between hyperuniform and uncorrelated patterns reducing $\Jc$ in the latter: (i)~the existence of large regions without pinning centers and (ii)~clusters of the defects located too close to each other. Indeed, large regions of pure superconductor allow vortices to move freely. On the other hand, two defects located too close to each other might not produce a strong enough potential barrier preventing a vortex jumping from one defect to the other. A less important effect is the higher probability of overlapping defects for uncorrelated placement, which effectively lowers the number of pinning sites.

\section{Conclusions} \label{sec:conclusions}

We carried out large scale time-dependent Ginzburg-Landau simulations of vortex dynamics in thin, hexagonally patterned superconducting films and reproduced experimentally measured critical currents in patterned MoGe thin-film samples. We studied the vortex dynamics inside the sample and revealed the underlying mechanisms for the critical current dependence on the magnetic field. In particular, we demonstrated how the position of peaks of the magnetic-field-dependent critical current is influenced by the depth and diameter of the individual defect, and we discussed the different types of vortex dynamics near these peaks. Overall, the field dependence of the critical current strongly depends on the defect morphology. Finally, we observed that spatial randomness in the position of inclusions smooths the critical current curve, i.e., increases or decreases it in certain value ranges. We found that a hyperuniform placement of inclusions can generate larger critical current than completely uncorrelated random pinscapes.

\paragraph*{Acknowledgments.}
We are delighted to thank A.~E.~Koshelev and G.~Kimmel for illuminating discussions. The computational work was supported by the Scientific Discovery through Advanced Computing (SciDAC) program funded by the U.~S. Department of Energy, Office of Science, Advanced Scientific Computing Research and Basic Energy Science. The computational part of this work was performed on Titan at the LCF at Oak Ridge National Laboratory (DOE Contract No. DE-AC05-00OR22725) and GAEA at Northern Illinois University. The experimental study at Argonne National Laboratory was supported by the U.~S. Department of Energy, Office of Science, Basic Energy Sciences, Materials Sciences and Engineering Division. The nanopatterning and morphological analysis were performed at Argonne National Laboratory's Center for Nanoscale Materials. Z.~L.~X. acknowledges NSF Grant No. DMR-1407175.

\bibliography{hexpinning}

\end{document}